\documentclass[
              ]{jetp} 

\newcommand{\be}{\begin{equation}}
\newcommand{\ee}{\end{equation}} 
\newcommand{\bea}{\begin{eqnarray}}
\newcommand{\eea}{\end{eqnarray}}

\begin{document}
\English

\title{Sweeping effect and Taylor's hypothesis  via correlation function
} 


 \setaffiliation1{Department of Physics, Indian Institute of Technology, Kanpur 208016, India }
 \setaffiliation2{Centre for Fluid and Complex Systems, Coventry University, Coventry CV1 5FB, UK} 
 \setauthor{Mahendra}{Kumar~Verma}{1} 
\email{mkv@iitk.ac.in}
 \setauthor{Abhishek}{Kumar}{2} 
\email{abhishek.kir@gmail.com }
 \setauthor{Akanksha}{Gupta}{1} 
\email{akgupt@iitk.ac.in}

\abstract{We demonstrate the sweeping effect in turbulence using numerical simulations of hydrodynamic turbulence  without a mean velocity. The velocity correlation function, $C({\bf k},\tau)$, decays with time due to the  eddy viscosity.  In addition, $C({\bf k},\tau)$  shows oscillations due to the sweeping effect by ``random mean velocity field$"$ ${\bf \tilde{U}}_0$.  We also perform numerical simulation with a mean velocity ${\bf U}_0= 10\hat{z}$  for which  $C({\bf k},\tau)$ exhibits damped oscillations with the frequency of $|{\bf U}_0| k$ and decay time scale corresponding to the ${\bf U}_0=0$ case.  For ${\bf U}_0=10\hat{z}$, the phase of $C({\bf k},\tau)$ show the sweeping effect, but it is overshadowed by oscillations caused by ${\bf U}_0$.  We also demonstrate that for ${\bf U}_0=0$ and $10\hat{z}$, the frequency spectra of the velocity fields measured by real-space probes are respectively $f^{-2}$ and $f^{-5/3}$; these spectra are related to the Lagrangian and Eulerian space-time correlations respectively. 
}

\maketitle


\section{Introduction}
The incompressible Navier--Stokes equations of a flow that is moving with a mean velocity of ${\mathbf U}_0$ is
\begin{eqnarray}
\frac{ \partial{\bf u}}{\partial t} + ( {\bf u} \cdot \nabla) {\bf u}+ ({\mathbf U}_0 \cdot \nabla) {\bf u}& = &  -\nabla p + \nu \nabla^2  {\bf u} + {\bf f}, 
\label{eq:NS1} \\
\nabla \cdot  {\bf u}  & = & 0, \label{eq:NS2}
\end{eqnarray}
where ${\bf u}$ is the velocity fluctuation with a zero mean, ${\bf f}$ is the external force,  $p$ is the pressure, and $\nu$ is the kinematic viscosity.   One of the  important principles of classical physics is Galilean invariance, according to which the laws of physics are the same in all inertial frames (frames moving with constant velocities  relative to each other). Naturally, the Navier--Stokes equations, which are essentially  Newton's laws for fluid flows, exhibits this symmetry~\cite{Lesieur:book:Turbulence,Frisch:book,Davidson:book:Turbulence,McComb:book:Turbulence,McComb:book:new}.   As a consequence of this symmetry, the flow properties of the fluid in the laboratory reference frame (in which the fluid moves with a mean velocity of ${\mathbf U}_0$) and in the co-moving reference frame  (${\mathbf U}_0=0$) are the same.

The velocity field of a turbulent flow is random, hence it is typically characterised by its correlations. There have been several major advances in the understanding the correlations in homogeneous and isotropic turbulence, most notably by Kolmogorov~\cite{Kolmogorov:DANS1941Structure,Kolmogorov:DANS1941Dissipation} who showed that in the inertial range, the velocity correlation $C({\bf k}) =  \langle |\mathbf u(\mathbf k)|^2 \rangle = K_\mathrm{Ko} \Pi^{2/3} k^{-5/3}/(4\pi k^2)$, where $\Pi$ is the energy flux, and $K_\mathrm{Ko}$ is the Kolmogorov constant.  The corresponding one-dimensional energy spectrum is $ E(k) = K_\mathrm{Ko} \Pi^{2/3} k^{-5/3}$.

Kraichnan (1964) \cite{Kraichnan:PF1964Eulerian}  argued that in the presence of a ``random mean velocity$"$ field, ${\tilde{\mathbf U}}_0$, Eulerian field theory does not yield  Kolmogorov's spectrum.  In particular, Kraichnan (1964) considered a  fluid flow with ${\tilde{\mathbf U}}_0$ that is constant in space and time but has a Gaussian and isotropic distribution over an ensemble of realisations.  Then he employed direct interaction approximation (DIA) to close the hierarchy of equations and showed that $E(k) \sim (\Pi \tilde{U}_\mathrm{0})^{1/2} k^{-3/2}$, where $\tilde{U}_\mathrm{0}$ is the { root mean square (rms)}  value of the mean velocity.    Kraichnan (1964) argued that the above deviation of the energy spectrum from the experimentally observed Kolmogorov's $k^{-5/3}$ energy spectrum is due to the sweeping of  small-scale fluid structures by the large energy-containing eddies.  This phenomenon is called {\em sweeping effect}.   Based on the above observations, Kraichnan (1964)   emphasised that the Eulerian formalism is inadequate for obtaining Kolmogorov's spectrum for a fully developed fluid turbulence. Later, he developed Lagrangian field theory for hydrodynamic turbulence that is consistent with the Kolmogorov's 5/3 theory of turbulence ~(see Kraichnan 1965, and other related papers). The above framework is called  {\em random Galilean invariance}. 

 There have been several attempts to test the sweeping effect.  Kraichnan (1964)  had argued that the nonlinear  time scale  is $1/(k \tilde{U}_0)$ due to the dependence on the mean random velocity $\tilde{U}_0$, and hence the energy spectrum $E(k) \sim (\Pi \tilde{U}_0)^{1/2} k^{-3/2}$.   Sanada $\&$
Shanmugasundaram (1992) \cite{Sanada:PF1992} computed the time scale for the decay of the correlation function for various $k$'s, and argued it to vary as $k^{-1}$, in line with the predictions of Kraichnan (1964).   Based on these results, Sanada $\&$ Shanmugasundaram (1992)  argued that their correlation function validates the sweeping effect. { In a related work, Drivas \textit{et al. }(2017) \cite{Drivas:PRF2017} employed  spatial filtering  to study the sweeping effect  on small-scale velocities by a large-scale flow. They showed consistency between results of direct numerical simulation and large-eddy simulation with appropriate filtering. }

He \textit{et al.}~(2010) \cite{He:PRE2010} and He $\&$ Tong~(2011) \cite{He:PRE2011} proposed  elliptic model in which the isocorrelation lines of  two-point two-time velocity correlations are ellipses parametrised by the mean and sweeping velocities.  Note that the mean velocity ${\bf U}_0$ is related to the Taylor's frozen-in hypothesis (to be described below).  Researchers observed that the elliptic model describes several  experimental and direct numerical simulation (DNS) data more accurately than the classic Taylor hypothesis.  Thus, the elliptic model  validates the sweeping effect in hydrodynamic turbulence.  Wilczek $\&$ Narita (2012) \cite{Wilczek:PRE2012} derived the frequency spectrum of hydrodynamic turbulence based on sweeping effect and a mean flow.  Their results are consistent with the sweeping effect and elliptic model. 

A related phenomenon is Taylor's hypothesis of frozen-in turbulence.  Taylor (1938) \cite{Taylor:PRS1938} proposed that the velocity measurement at a point in a fully-developed turbulent flow moving with a constant velocity ${\bf U}_0$    (e.g. in a wind tunnel)  can be used to study the velocity correlations.  This is because the mean flow  advects the {\em frozen-in} fluctuations, and the stationary probe in the fluid measures the fluctuations along a line.  Here, the frequency spectrum of the measured time series is expected to show $f^{-5/3}$, where $f$ is the  frequency.   { This proposal, {\em Taylor's frozen-in turbulence hypothesis}, has been used in many experiments to ascertain Kolmogorov's spectrum~\cite{Tennekes:book}. }  As discussed above, Taylor's frozen-in hypothesis is incorporated in the elliptic model and in Wilczek $\&$ Narita (2012)'s model \cite{Wilczek:PRE2012}.  

 In this paper, our approach is somewhat different from  earlier ones.  We compute the normalized correlation function and find this to be complex, unlike Sanada $\&$ Shanmugasundaram (1992)'s function which is real.  The phase of the correlation function helps us deduce the random mean velocity that is responsible for the sweeping effect.  We thus provide a definitive evidence for the sweeping effect.  In addition, we also analyse the correlation function with and without a mean velocity, as well as the frequency spectrum.   In \S\ref{sec:Taylor's_hypothesis} we show that the frequency spectrum $E(f)\sim f^{-2}$ for turbulent flow in the absence of a constant mean velocity field ${\bf U}_0$, and $E(f)\sim f^{-5/3}$ for large $ U_0 $.

In the next two sections, we briefly describe the Green's functions, correlation function, and sweeping effect in hydrodynamic turbulence.  In \S\ref{sec:sweeping_effect_numerical} we demonstrate the signature of sweeping effect using numerical simulation. In  \S\ref{sec:Taylor's_hypothesis}  we describe the frequency spectra of hydrodynamic turbulence in the absence and presence  of a mean flow. We conclude in \S\ref{sec:conclusion}.

\section{A brief review of Green's function and correlation function in hydrodynamic turbulence}
\label{sec:Green's_fn}

Kraichnan (1964) derived the  sweeping effect using {\em direct interaction approximation (DIA)}~(Kraichnan 1959) \cite{Kraichnan:JFM1959}.  We will sketch sweeping effect in the next section.  However its description requires some terminologies, such as Green's function, correlation function, and effective viscosity, which will be briefly described below. See Kraichnan (1959) \cite{Kraichnan:JFM1959} and Leslie (1973) \cite{Leslie:book} for details.
 
A  linearised version of Eq.~(\ref{eq:NS1}) in  Fourier space is
\be
\left( \frac{ \partial}{\partial t}  +i  {\mathbf U_0   \cdot \mathbf k}  +   \nu k^{2} \right) {\bf u}(\mathbf{{k}})   =  {\bf f}(\mathbf{k}),  \label{eq:uk_dot}  
\ee
where ${\bf k}$ is the wavevector,  The corresponding equation for the Green's function is
\bea
\left( \frac{ \partial}{\partial t}  +i  {\mathbf U_0   \cdot \mathbf k}  +   \nu k^{2} \right)   G({\bf k},t,t')  & = & \delta(t-t'),
\label{eq:Green_linear}
\eea
whose solution is
\be
G({\bf k},\tau) =\theta(\tau) \exp{(i {\mathbf U_0   \cdot \mathbf k} t)}  \exp{(- \nu k^2 \tau)},
\label{eq:Gkt_nu0}
\ee
where $\tau = t-t'$, and $\theta(\tau)$ is the step function.  

In addition, the equal-time correction function, $C({\bf k},0)$, and unequal time correction function, $C({\bf k},\tau)$, are defined as
\bea
C({\bf k},0) & = & \langle |\mathbf u(\mathbf k, t)|^2 \rangle, \\
C({\bf k},\tau) & = & \langle \mathbf u(\mathbf k, t)\cdot \mathbf u^{*}(\mathbf k, t+\tau) \rangle.
\label{eq:C_k_tau_def}
\eea
In the above, the averaging could be either ensemble or temporal (due to homogeneity in time). The  ratio of the two correlation function is the  normalised  correlation function:
\begin{equation}
R(\mathbf k, \tau) = \frac{C(\mathbf k, \tau)}{C(\mathbf k, 0)}.
  \label{eq:R}
\end{equation} 
A   generalisation of fluctuation-dissipation theorem to hydrodynamics yields~\cite{Kiyani:PRE2004}
\be
R(\mathbf k, \tau) = G({\bf k},\tau) =\theta(\tau) \exp{(i {\mathbf U_0   \cdot \mathbf k} t)}  \exp{(- \nu k^2 \tau)}.
\label{eq:Rkt_linear}
\ee
That is, the normalised correlation function exhibits damped oscillations---oscillations due to ${\bf U}_0$, while damping arising from the viscous part.

Researchers attempted to incorporate the effects of nonlinearity in the above functions.  The methods used are DIA~\cite{Kraichnan:JFM1959, Leslie:book}, Lagrangian field theory~\cite{Kraichnan:PF1965Lagrangian_history}, renormalisation groups \cite{Yakhot:JSC1986, McComb:book:Turbulence, DeDominicis:PRA1979, Zhou:PR2010}, etc.  We do not detail these methods here, but we state several important results derived using these computations:
\begin{enumerate}
\item Using field theory and certain assumptions, researchers have been able to show that the nonlinearity yields  enhanced viscosity at a wavenumber $k$ in the following manner:
\be
\nu \rightarrow \nu + \nu(k),
\label{eq:total_visc}
\ee
where $\nu(k)$,  called ``effective viscosity$"$ or ``renormalized viscosity$"$, is
\be
\nu(k) = \nu_* \sqrt{K_\mathrm{Ko}} \epsilon^{1/3} k^{-4/3},
\label{eq:nu_k}
\ee
with $\nu_*$ as  a constant. {  Physically, $\nu(k)$ represents the effective viscosity at wavenumber $k$.  For large $k$'s (in the inertial range), $\nu(k) \gg \nu$, hence the total viscosity is essentially $\nu(k)$.  This  viscosity leads to enhanced mixing.}  In other words, the effective Navier-Stokes equation in the presence of nonlinearity is
\be
\left( \frac{ \partial}{\partial t}  +i  {\mathbf U_0   \cdot \mathbf k}  +  [\nu +  \nu(k)] k^{2} \right) {\bf u}(\mathbf{{k}})   = {\bf N}(\mathbf{k})  +  {\bf f}(\mathbf{k}),  \label{eq:uk_dot_nlin}  
\ee
where ${\bf N}(\mathbf{k})$ is the nonlinear term (including the pressure gradient).  Refer to \cite{Yakhot:JSC1986, McComb:book:Turbulence, DeDominicis:PRA1979, Zhou:PR2010} for details. Also see Appendix A. 

\item Using Eq.~(\ref{eq:uk_dot_nlin}) and  certain assumptions on the perturbation, Green's function of Eq.~(\ref{eq:Gkt_nu0}) gets transformed to the following form for the Navier-Stokes equations with the nonlinear terms: 
\bea
G({\bf k},\tau) & = &\theta(\tau) \exp{(i {\mathbf U_0   \cdot \mathbf k} t)}  \exp{(- \nu(k) k^2 \tau)}
\nonumber \\
& = & \theta(\tau) \exp{(i {\mathbf U_0   \cdot \mathbf k} t)}  \exp{(-\tau/\tau_c)},  \label{eq:Gkt_nuk}  
\eea
where  
\be
\tau_c = \frac{1}{\nu(k) k^2} \sim \frac{1}{\epsilon^{1/3} k^{2/3}}.
\label{eq:tauc_Euler}
\ee 
is the decay time scale.  Since $\nu(k) \gg \nu$, the decay time scale for Eq.~(\ref{eq:Gkt_nuk}) is much smaller than the corresponding time scale for Eq.~(\ref{eq:Gkt_nu0}).  The above Green's function is called ``dressed Green's function$"$ in field theory.   

\item For the nonlinear equation, using field-theoretic treatment and  generalisation of fluctuation-dissipation theorem,  the correlation function of Eq.~(\ref{eq:Rkt_linear}) is generalised to 
\be
R({\bf k},\tau) = G({\bf k},\tau)  =  \theta(\tau) \exp{(i {\mathbf U_0   \cdot \mathbf k} t)}  \exp{(- \nu(k) k^2 \tau)}.
\label{eq:Rkt_nonlinear}
\ee
That is, the decay time scale for the correlation function is same as that for the Green's function.

\item In the absence of ${\bf U}_0$, the correlation and Green's functions are:
\be
R({\bf k},\tau) = G({\bf k},\tau)  =  \theta(\tau)   \exp{(- \nu(k) k^2 \tau)}.
\ee
The above function exhibits pure damping.

\end{enumerate}

 In the following section we provide a brief introduction  to the sweeping effect.

\section{Brief description of sweeping effect}
\label{sec:sweeping_effect}

In this section, we briefly describe the {\em sweeping effect} \cite{Kraichnan:PF1964Eulerian}.   Kraichnan assumed that the velocity fluctuations of Navier--Stokes equations is advected by  {\em random large-scale flow}, $\tilde{\bf U}_0$.  For this case, Kraichnan (1964) ignored the viscous and nonlinear terms, and simplified Eq.~(\ref{eq:uk_dot_nlin}) to 
\be
\left( \frac{ \partial}{\partial t}  +i  \tilde{\bf U}_0   \cdot \mathbf{k}    \right) {\bf u}(\mathbf{{k}})   =   {\bf f}(\mathbf{k}). \label{eq:uk_dot_Kriachnan}  
\ee
For the above equation, the normalized correlation function is obtained by setting $\nu=0$ in Eq.~(\ref{eq:Rkt_linear}):
\be
R({\bf k},\tau) =  \theta(\tau) \exp{(i  \tilde{\bf U}_0   \cdot \mathbf{k}  t)}.
\ee
Kraichnan (1964) further assumed that $\tilde{\bf U}_0$ is constant in time, but it is spatially varying with a Gaussian distribution.  Under these assumptions, the averaged correlation function has the following form   (also see Wilczek $\&$ Narita 2012  \cite{Wilczek:PRE2012}):
 \bea
R({\bf k}, \tau) & = &    \langle \exp[i {\bf k} \cdot \tilde{\bf U}_0 \tau]  \rangle = \exp\left[ - \frac{\langle \tilde{U}_0^2 \rangle k^2 \tau^2}{6} \right].  
 \label{eq:R_Kraichnan}
 \eea 
 { Note that the above averaging with gaussian $\tilde{U}_0$ differs from ensemble or temporal averaging performed for the correlation function of Eq.~(\ref{eq:C_k_tau_def}).}
 
  Using the above equation and field-theoretic arguments,  Kraichnan (1964) argued that the relevant nonlinear time scale is  $1/(k \tilde{U}_0)$ that would yield the following energy spectrum: 
  \be
  E(k) \sim (\Pi \tilde{U}_0)^{1/2} k^{-3/2}.
  \label{eq:Ek_3/3}
  \ee
    The above spectrum is very different from Kolmogorov's prediction that $E(k) \sim \Pi^{2/3} k^{-5/3}$, which is observed in experiments.  Based on these contradictions,  Kraichnan (1964) argued that Eulerian field theory is inadequate to reproduce $k^{-5/3}$ energy spectrum, and hence unsuitable for describing turbulence. He went on to develop Lagrangian field theory to reproduce the consistent energy spectrum~\cite{Kraichnan:PF1965Lagrangian_history}.  
  
In this paper we test the sweeping effect using numerical simulation.  Note that due random nature of large-scale flow $\tilde{\bf U}_0$,  
\bea
R(\mathbf k, \tau) & = & \exp(-\tau/\tau_c(k)) \exp(i \tilde{\bf U}_0 \cdot {\bf k} \tau) \nonumber \\
& \rightarrow &  \exp(-\tau/\tau_c(k)) \exp(i  c k U_0 \tau),
 \label{eq:Rk_tau_with_U_eq_0}
\eea
where $c$ is a random number that can take both positive and negative values. In the present paper we compute $R({\bf k},\tau) $ and  look for a signature of random $\tilde{U}_0$ in the phase of $R(\mathbf k, \tau)$.  A nonzero phase in Eq.~(\ref{eq:Rk_tau_with_U_eq_0}) would signal a presence of $\tilde{\bf U}_0$, hence the sweeping effect.  { Note that our proposed correlation function of Eq.~(\ref{eq:Rk_tau_with_U0}) differs from Eq.~(\ref{eq:R_Kraichnan})  of Kraichnan (1964). We do not make any assumption on the probability distribution of $ \tilde{\bf U}_0$.     This process helps us examine oscillations in $R(\mathbf k, \tau)$  induced by $\tilde{\bf U}_0$. }

In the presence of a mean velocity field ${\bf U}_0$, the correlation function of Eq.~(\ref{eq:Rkt_nonlinear}) with sweeping effect is expected to be of the following form:
\bea
R(\mathbf k, \tau) & = & \exp(-\tau/\tau_c(k)) \exp(i {\bf U}_0 \cdot {\bf k} \tau + i \tilde{\bf U}_0 \cdot {\bf k} \tau) \nonumber \\
& \rightarrow &  \exp(-\tau/\tau_c(k)) \exp(i  c k U_0 \tau) \exp(i {\bf U}_0 \cdot {\bf k} \tau). 
 \label{eq:Rk_tau_with_U0}
\eea
 The above correlation function includes sweeping effect along with oscillations arising due to  ${\bf U}_0$.  The Fourier transfer of the above equation to the frequency space yields the following Green's function in ${\bf k}, \omega$ space:
\be
G({\bf k},\omega) = \frac{1}{-i \omega + \nu(k) k^2 + i {\bf U}_0 \cdot {\bf k} +  i c k  \tilde{U}_0(k) }.
\label{eq:G_k_omega_with_U0} 
\ee

 In the next section we provide numerical evidences for the sweeping effect.
   
\section{New evidences for the sweeping effect}
\label{sec:sweeping_effect_numerical}

 In this section,  we demonstrate existence of wavenumber dependent phases of $R(\mathbf k, \tau) $, thus signalling presence of sweeping effect.  We perform numerical simulation of  Navier--Stokes equations in the turbulent regime for the mean velocity $\mathbf U_0=0$.  We employ pseudospectral code TARANG~\cite{Chatterjee:JPDC17}  to simulate the flow  on  $512^3$  and $1024^3$ grids with random forcing.  { For forcing, we employ the procedure proposed by Carati \textit{et al.} (1995) \cite{Carati:PF1995}.}
 We use the fourth-order Runge Kutta (RK4) scheme for  time stepping, 2/3 rule for dealiasing, and CFL condition for computing $\Delta t$.  The Reynolds number of the runs  are $u_\mathrm{rms} L/\nu =5.7 \times 10^3$  for $512^3$ grid, and $1.3 \times 10^4$ for $1024^3$ grid, where $u_\mathrm{rms}$ is the rms value of the velocity fluctuations, and $L$ is the box size. For the present set of simulations, we use $L=2\pi$.  { The unit of time for our simulation is eddy turnover time, $L/u_\mathrm{rms}$.}  The parameters of our simulations  are described in Table~\ref{table:simulation_details}. All our simulations are fully resolved since $k_{\rm max}\eta >1$, where $k_{\rm max}$ is the maximum wave number of the run, and $\eta$ is the Kolmogorov length scale.

\begin{figure}
\begin{center}
\includegraphics[scale = 0.7]{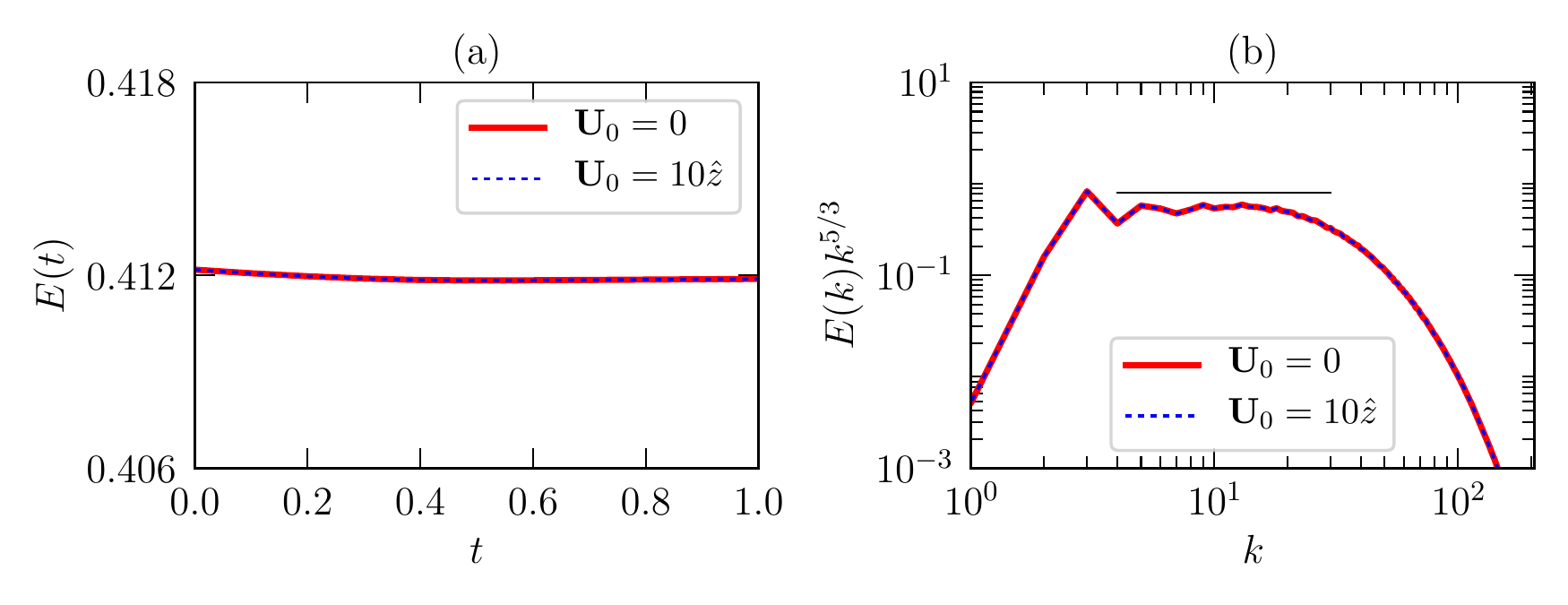}
\end{center}
\caption{For ${\bf U}_0=0$ and ${\bf U}_0=10\hat{z}$, (a) plots of total energy of the velocity fluctuation vs.~$t$ (in units of eddy turnover time). (b) Plots of the normalized kinetic energy spectrum $E(k)k^{5/3}$ vs.~$k$.    Here $E(t)$ and $E(k)$ are identical for ${\bf U}_0=0$ and $10\hat{z}$   due to the Galilean invariance of the fluid equations.}
\label{fig:energy_spectrum}
\end{figure}

We evolve the flow with ${\bf U}_0=0$ till a steady state is reached.  At this point, we fork the above run to new two simulations with ${\mathbf U}_0 = 0$ and ${\mathbf U}_0=10 \hat{z}$.  The new runs are carried up to one eddy turnover time each.  For  ${\bf U}_0=0$ and $10\hat{z}$,  the  temporal evolution of the fluctuating energy, as well as the energy spectra, are identical, as illustrated in Fig.~\ref{fig:energy_spectrum}; this is consistent with the Galilean invariance of the  Navier--Stokes equations.    These results, however, are based on equal-time correlations;  subtleties however emerge when we study the temporal correlations of the velocity Fourier modes.

\begin{figure}
\begin{center}
\includegraphics[scale = 0.7]{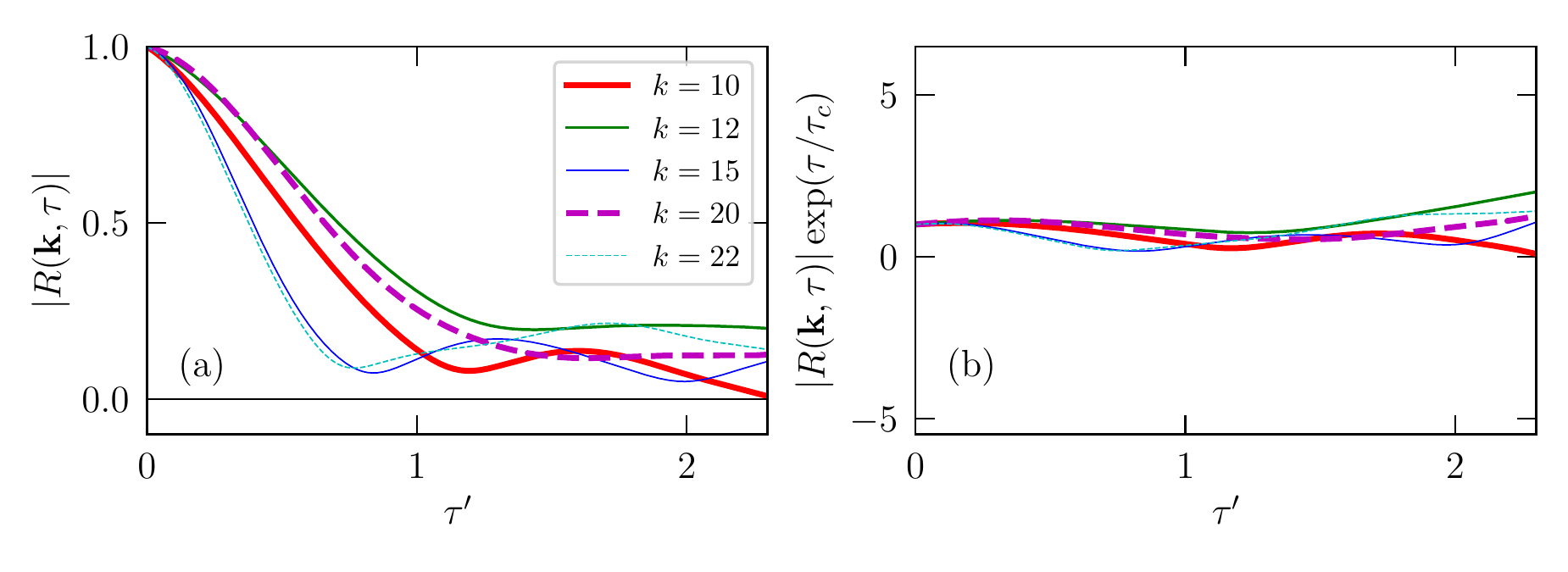}
\end{center}
\caption{For $\mathbf U_0 = 0$ and  $k = 10,12,15,20,22$ (inertial range wavenumbers), (a) plots of the absolute value of normalised correlation function, $|R(\mathbf k, \tau)|$ vs.~$\tau'=\tau/\tau_c$. It decays exponentially in time  as in Eq.~(\ref{eq:Rk_tau_with_U_eq_0}). (b)Plots of $|R(\mathbf k, \tau)| \exp(\tau/\tau_c)$ vs.~$\tau'$, which is approximately flat.}
\label{fig:corr}
\end{figure}

 Using the steady state numerical data of $1024^3$ grid, we compute the normalised  correlation function $R(\mathbf k, \tau)$ of Eq.~(\ref{eq:R}) for  $k = 10,12,15,20$, and 22 that lie in the inertial range.   { The correlation  $R(\mathbf k, \tau)$ was time overaged over  $12500$ data points collected over $1.3$ eddy turnover time.}  We observe that   $R(\mathbf k, \tau)$ is complex, thus providing clues for the  sweeping effect in the flow.  In Fig.~\ref{fig:corr}(a) we plot $|R(\mathbf k, \tau)|$ that  decays  exponentially with time with an approximate time scale of $\tau_c(k)$ given by Eq.~(\ref{eq:tauc_Euler}).  For validation of this conjecture, in  Fig.~\ref{fig:corr}(b) we plot  $|R(\mathbf k, \tau)| \exp(\tau/\tau_c)$ which are  approximate flat curves for all $k$'s.

\begin{figure}
\begin{center}
\includegraphics[scale = 0.7]{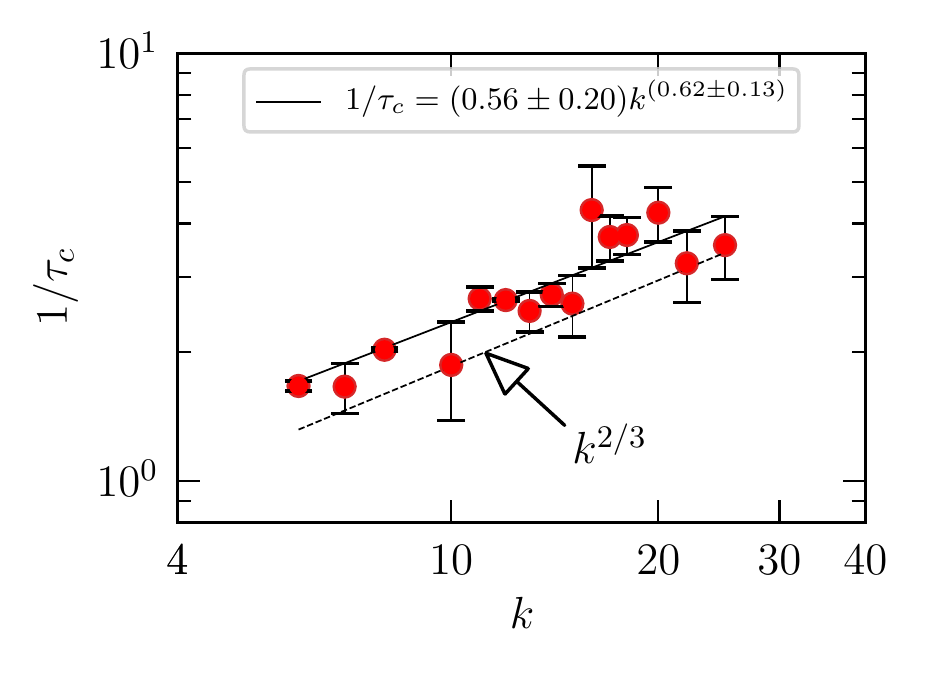}
\end{center}
\caption{Plot  of $\tau_c^{-1} $ vs.~$k$.   We observe that  $\tau_c^{-1} \sim  k^{0.62 \pm 0.13}$. The exponent being closer to 2/3 indicates that  Eq.~(\ref{eq:tauc_Euler}) provides a fair description of the decaying time scale.}
\label{fig:tau_c_k}
\end{figure}

We compute  $\tau_c(k)$ from the slope of $|R(\mathbf k, \tau)|$  in a semiology plot for a range of $k$'s.  A regression analysis of the data yields
\be
\tau_c(k)  \sim k^{-0.62\pm 0.13},
\ee
{for $k$ ranging from 6 to 25.} The slope of $-0.62$ is consistent with the predicted $-2/3$ slope of Eq.~(\ref{eq:tauc_Euler}).   We exhibit the plot in Fig.~\ref{fig:tau_c_k} that exhibits some scatter, which is possibly due to the random velocity field as postulated in the sweeping effect.  This observation is contrary to that of  Sanada $\&$ Shanmugasundaram (1992) who argued that $\tau_c \sim k^{-1}$ { based on Kraichnan (1964)'s sweeting effect arguments according to which $\tau_c \sim 1/(k \tilde{U}_0)$.   }

\begin{figure}
\begin{center}
\includegraphics[scale = 0.7]{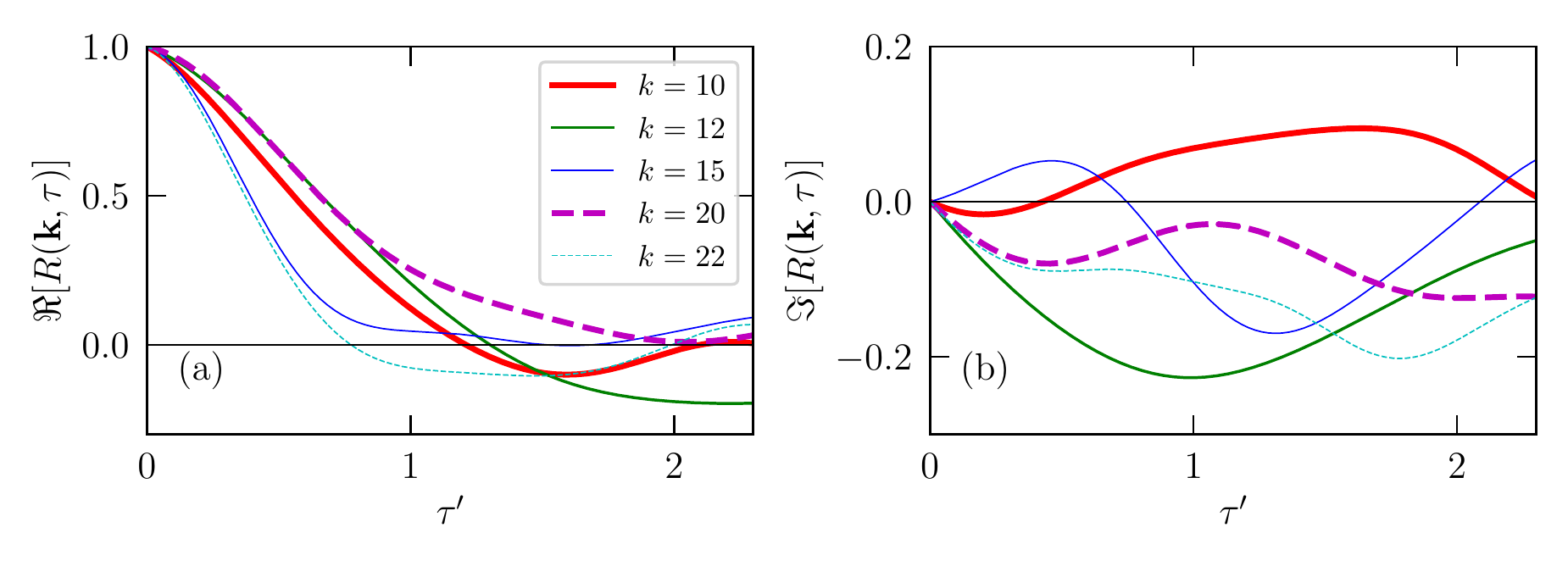}
\includegraphics[scale = 0.7]{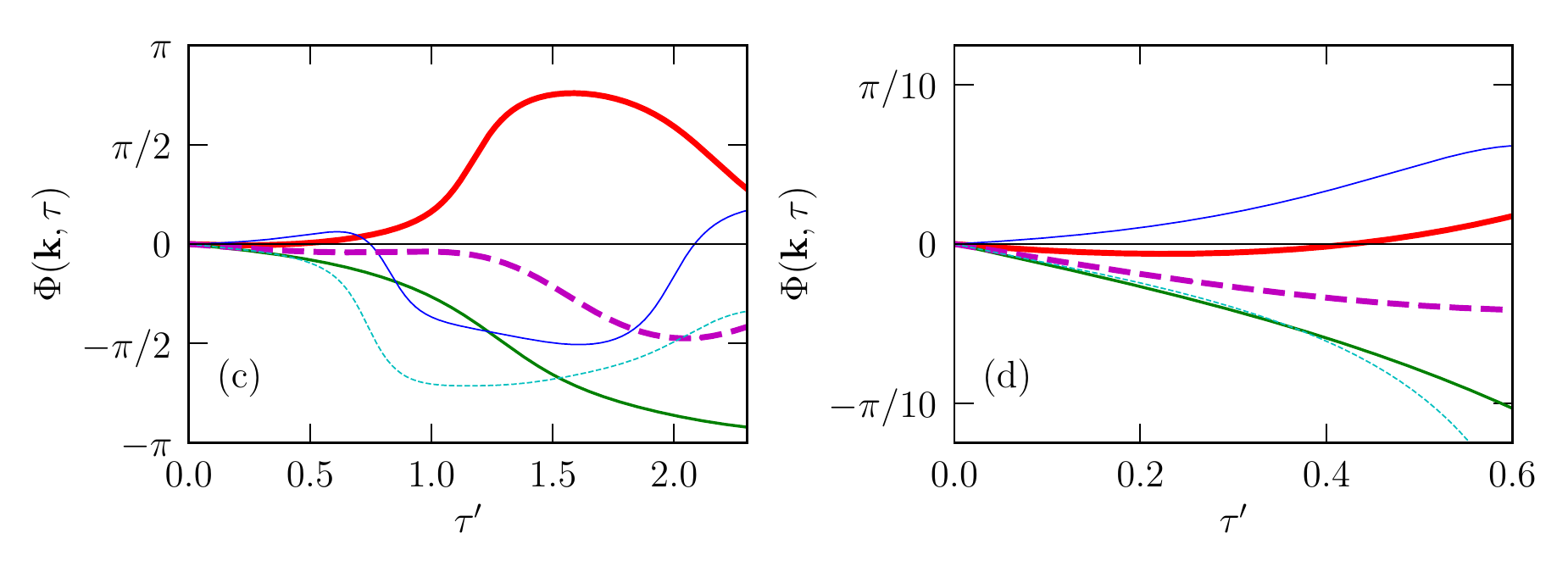}
\end{center}
\caption{For $\mathbf U_0 = 0$ and  $k = 10,12,15,20,22$ (inertial range wavenumbers), plots of the $\Re[R(\mathbf k, \tau)]$,  (b) $\Im[R(\mathbf k, \tau)]$, and (c,d) $\Phi(\mathbf k, \tau)$, where $R(\mathbf k, \tau)$ is the normalized correlation function.  Subfigure (d) is a zoomed view of (c) for $\tau'=0:0.6$.  The real part exhibits decaying oscillations, while the imaginary part shows oscillations, consistent with Eq.~(\ref{eq:Rk_tau_with_U_eq_0}).  The phases for various ${\bf k}$'s exhibit monotonic increase with time till $\tau' = \tau_2 = 0.6$ due to $\tilde{{\bf U}}_0$, thus demonstrating the sweeping effect.}
\label{fig:corr2}
\end{figure}

In Fig.~\ref{fig:corr2} we plot   $\Re[R(\mathbf k, \tau)]$, $\Im[R(\mathbf k, \tau)]$, and the phase of $R(\mathbf k, \tau)$, which is defined as
\begin{equation}
\Phi(k,\tau) = \tan^{-1} \frac{\Im[R(\mathbf k, \tau)]}{\Re[R(\mathbf k, \tau)]}.
\end{equation}
The phase $\Phi(k,\tau) $  varies linearly till $\tau_2 \approx 0.6 \tau_c$, which is the duration for the constancy of $\tilde{\bf U}_0$.  We can estimate  $\tilde{\bf U}_0$ from the phase using Eq.~(\ref{eq:Rk_tau_with_U_eq_0}) with $|c|=1$ (which is an assumption).  From Fig.~\ref{fig:corr2} we deduce the following properties for $\Phi(k,\tau) $:
\begin{enumerate}
\item  The phase $\Phi({\mathbf k},\tau)$  increases linearly with time till $\tau \approx \tau_2$, hence $\Phi({\mathbf k},\tau) \propto \tau$ till $\tau \approx \tau_2$.
\item  In Fig.~\ref{fig:corr2}, the slopes of the $\Phi({\mathbf k},\tau)$ for various $k$'s are different, hence $\Phi({\mathbf k},\tau) \ne D\tau$ with a constant $D$ for all $k$'s.  Therefore, we can easily conclude that the Fourier modes are not advected by a constant mean velocity field, say ${\bf U}_0$.
\item  The slopes of $\Phi({\mathbf k},\tau)$ come with both positive and negative signs, thus confirming random sweeping effect. 
\end{enumerate}
{ Thus, the nonzero phase $\Phi(k,\tau) $ provides evidence for the sweeping effect.  In addition, the real and imaginary parts of  $R(\mathbf k, \tau)$ contain the effects of both magnitude and phases, hence they show damped oscillations.  }

 Hence the numerical  correlation functions are consistent with Eq.~(\ref{eq:Rk_tau_with_U_eq_0}), thus providing a numerical demonstration of the {\em sweeping effect} proposed by Kraichnan (1964).  Physically, a Fourier mode ${\bf u}({\bf k})$ is being advected by the {\em random mean velocity field},   $\tilde{U}_0(\bf k)$.   The random velocity changes its direction and magnitude in around an eddy turnover time.  We observe that the phases are linear in $\tau$ only up to $\tau \approx \tau_2$.   The aforementioned wavenumber-dependent mean velocity field is in the similar spirit as the advection of eddies within eddies~\cite{Davidson:book:Turbulence,Pope:book,McComb:book:Turbulence}.     { It is important to note that the aforementioned time variation of $\tilde{U}_0(\bf k)$ is contrary to the assumption of Kraichnan (1964) that $\tilde{U}_0(\bf k)$  is constant in time. }
 A detailed analysis of $\tilde{\bf U}_0({\bf k})$  as a function of wavenumber and angles, as well as its probability distribution, will be performed in future.  

In the next section we analyse the sweeping effect in the presence of ${\bf U}_0$. 

\section{ Taylor's hypothesis for ${\bf U}_0 \ne 0$, and frequency spectrum}
\label{sec:Taylor's_hypothesis}

In the present section we compute $R(\mathbf k, \tau)$ for nonzero ${\bf U}_0$ using numerical data and compare it with Eq.~(\ref{eq:Rk_tau_with_U0}).  After that we will describe the frequency spectrum for zero and nonzero ${\bf U}_0$.

As argued in \S\ref{sec:Green's_fn}, for nonzero ${\bf U}_0$, the normalised correlation function  given by Eq.~(\ref{eq:Rk_tau_with_U0}).  Thus, the mean velocity field induces a factor of $ \exp(-i {\bf U}_0 \cdot {\bf k} \tau)$ in the correlation function in comparison to Eq.~(\ref{eq:Rk_tau_with_U_eq_0}) for ${\bf U}_0=0$.   To verify the above correlation function, we perform numerical simulation of Eqs.~(\ref{eq:NS1}, \ref{eq:NS2}) with ${\bf U}_0 = 10\hat{z}$ and compute  the correlation function $R(\mathbf k, \tau)$ for ${\bf k}=(0,0,10)$.
 
\begin{figure}
\begin{center}
\includegraphics[scale = 0.7]{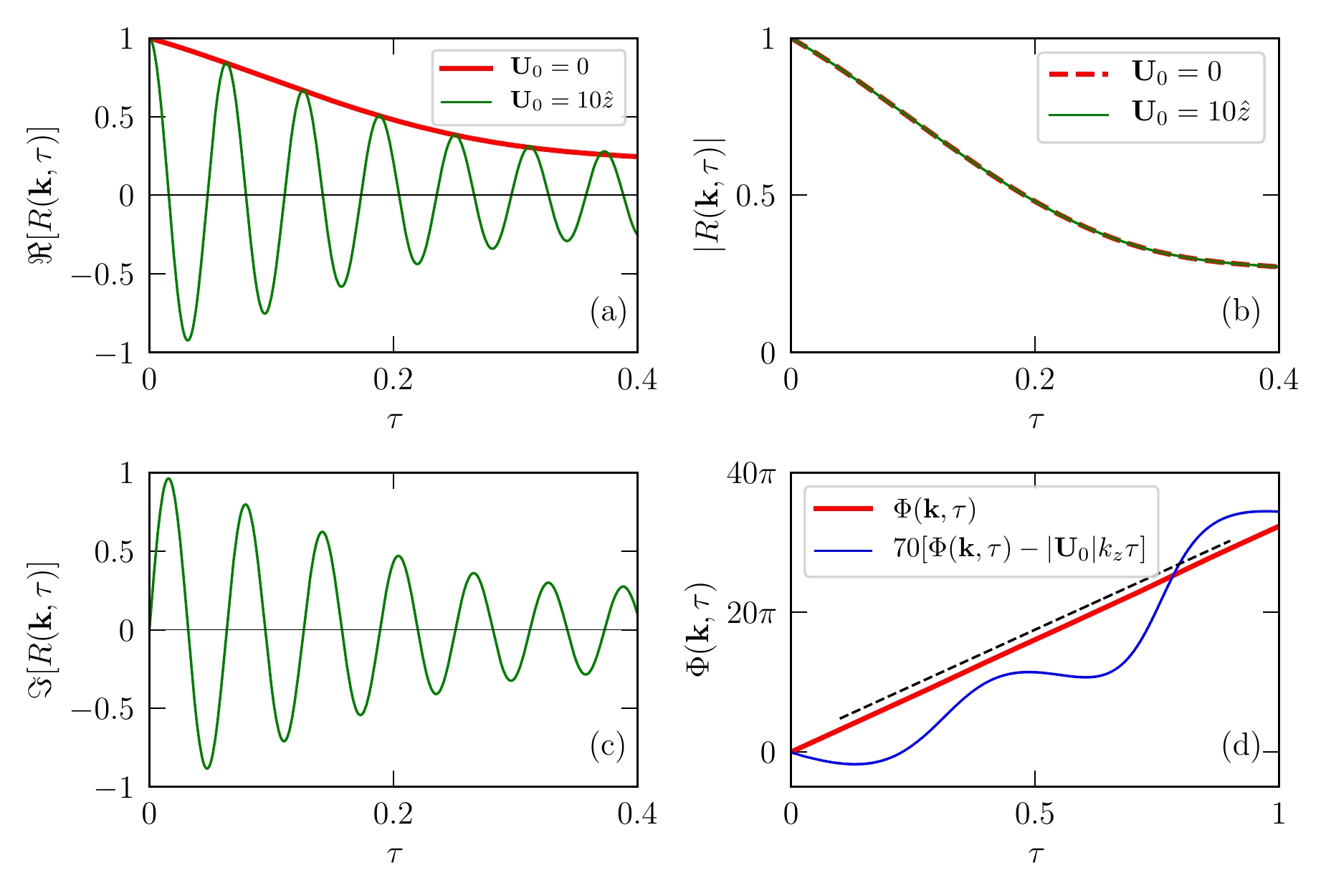}
\end{center}
\caption{For $\mathbf U_0 = 10$ and  $\mathbf k = (0,0,10)$ in the inertial range, plots of the normalised correlation function $R(\mathbf k, \tau)$ vs.~$\tau$:  (a) $\Re[R(\mathbf k, \tau)]$, (b) $|R(\mathbf k, \tau)|$, (c) $\Im[R(\mathbf k, \tau)]$, and (d) $\Phi(\mathbf k, \tau)$.  The real and imaginary parts exhibit damped oscillation with the frequency of $ |{\bf U}_0| k$ and damping time of $1/(\nu(k) k^2)$.  $|R(\mathbf k, \tau)|$ for ${\bf U}_0=0,10\hat{z}$ are identical, thus showing that the decay time scales for the two cases are the same; also, $|R(\mathbf k, \tau)|$ provides envelop to the real part.  The phase of  $R(\mathbf k, \tau)$ varies as $\Phi({\bf k},\tau) = |{\bf U}_0| k_z \tau + \delta$, where  $\delta$  arises due to the sweeping by the random large-scale flow structures.  The dashed black and  blue lines represent $|{\bf U}_0|  k_z \tau$ and $70 \delta$ (amplified by a factor for visualisation) respectively.  }
\label{fig:U10}
\end{figure}

 In Fig.~\ref{fig:U10}, we plot  the real and imaginary parts of the correlation $R(\mathbf k, \tau)$, as well as its magnitude and phase.  As shown in the figure, $|R(\mathbf k, \tau)|$ for ${\bf U}_0=0$ and $10\hat{z}$ are approximately the same.  However, both the real and imaginary parts of $R(\mathbf k, \tau)$ exhibit damped oscillations with a frequency of $\omega = k_z |{\bf U}_0|$ and a decay time scale of $1/(\nu(k) k^2)$. { The oscillations are due to the $\exp(-i {\bf U}_0 \cdot {\bf k} \tau)$  term.   Note that the damping time scales $\tau_c(k)$ are independent of ${\bf U}_0$, which is verified by the plot of Fig.~\ref{fig:U10}(a,b) in which the  envelops of $|R(\mathbf k, \tau)|$ and  $\Re[R(\mathbf k, \tau)]$ match  with  the corresponding plots for ${\bf U}_0=0$ (shown as red lines).}

 The correlation function also contains signatures of the {\em random sweeping effect} for ${\bf U}_0=10\hat{z}$.  In Fig.~\ref{fig:U10}(d), we plot the phase $\Phi$ of $R(\mathbf k, \tau)$, which is quite close to $U_0 k \tau$.  However, $\Phi - U_0 k \tau$ is nonzero, which is evident  from its magnified plot in Fig.~\ref{fig:U10}(d).  This deviation is due to the random sweeping effect by random mean field $\tilde{U}_0$. Thus, the small-scale fluctuations are swept by ${\bf U}_0=10 \hat{z}$ and by large-scale random velocity $\tilde{U}_0(k)$.  Thus, sweeping effect, though overshadowed by $U_0$, is present for nonzero $U_0$ as well.
 
{ In summary, for the normalised correlation function, the absolute value $|R(\mathbf k, \tau)|$   falls exponentially with a decay time scale of $\tau_c$, while the phase $\Phi(\mathbf k, \tau) \propto {\bf U}_0 \cdot {\bf k}+ c \tilde{U}_0 k$  that contains contributions from the mean velocity as well as from the random mean velocity (sweeping effect).  Note that $|R(\mathbf k, \tau)|$  is independent of $U_0$, but the real and imaginary parts of  $R(\mathbf k, \tau)$ contain the effects of both magnitude and phases, hence they exhibit  damped oscillations.}

We can use the spectral correlation function of Eq.~(\ref{eq:Rk_tau_with_U0}) to compute the the following spatio-temporal correlation~\cite{Wilczek:PRE2012}:
\bea
 C({\bf r}, \tau) & \sim & \int d{\bf k} C({\bf k}) \exp(-\nu(k) k^2 \tau - i{\bf U_0 \cdot k} \tau) \langle \exp(-i c k \tilde{U}_0( k)\tau)  \rangle \exp(i {\bf k} \cdot {\bf r}) \nonumber \\
 & = & \int d{\bf k} C({\bf k}) \exp(-\tau/\tau_c - i{\bf U_0 \cdot k} \tau )  \exp(-k^2 [\tilde{U}_0( k)]^2 \tau^2)  \exp(i {\bf k} \cdot {\bf r}).
\eea 
We obtain temporal correlation  $C(\tau)$ measured at the same location by setting ${\bf r}=0$.    Fourier transform of $C(\tau)$ yields the frequency spectrum $E(f)$, which is often measured in experiments.    Researchers have exploited the above hypothesis to measure turbulence spectrum in many fluid and plasma experiments, for example in wind tunnels (Tennekes $\&$ Lumley 1972 \cite{Tennekes:book}), and in the solar wind  using extraterrestrial spacecrafts (Matthaeus $\&$ Goldstein 1982 \cite{Matthaeus:JGR1982rugged}).

\begin{figure}
\begin{center}
\includegraphics[scale = 0.7]{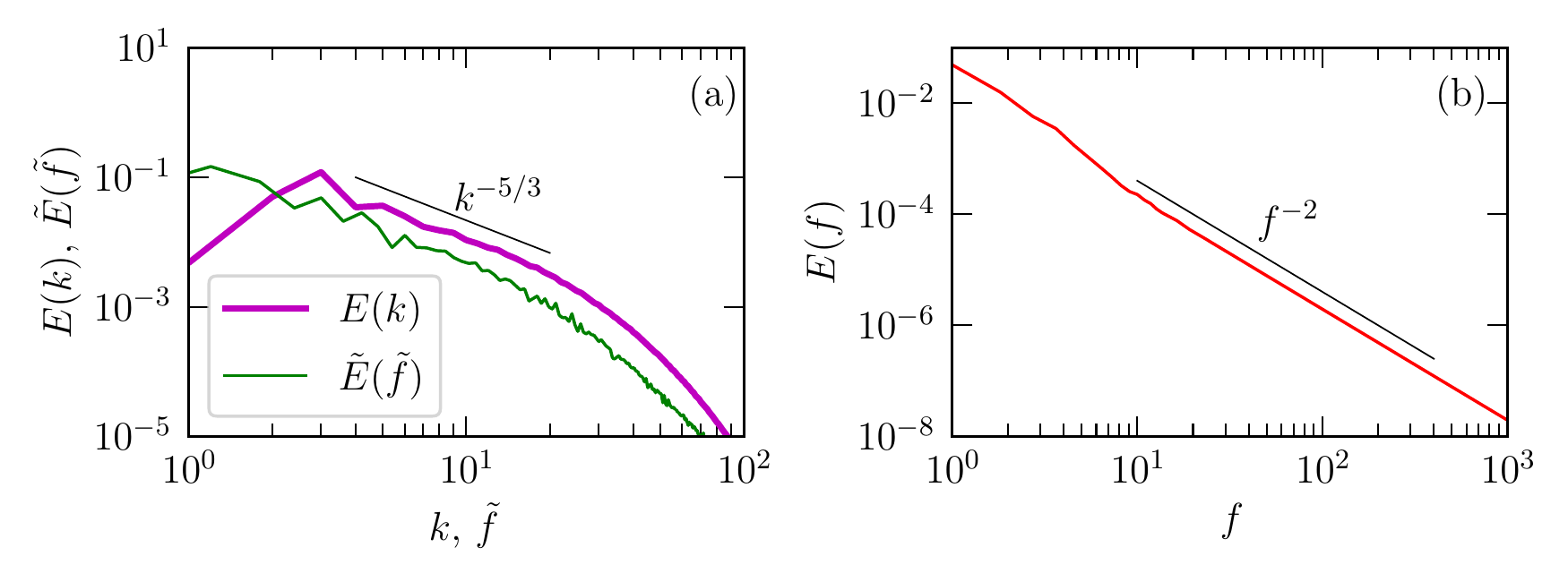}
\end{center}
\caption{(a) For ${\bf U}_0=10\hat{z}$, plots of the wavenumber spectrum $E(k)$ and the scaled frequency spectrum $E(f)$ for the velocity time series measured by  real-space probes.  The plot is averaged over 50 real-space probes located at random locations.  Here $\tilde{f} = f (2\pi)/U_0$ and $\tilde{E}(\tilde{f}) = E(f) U_0/(2\pi)$.  $E(f) \sim f^{-5/3}$, consistent with Taylor's frozen-in turbulence hypothesis.  (b) For  ${\bf U}_0=0$,  $E(f) \sim f^{-2}$, consistent with the sweeping effect.}
\label{fig:freq_spec}
\end{figure}

In Appendix~\ref{app:frequency_spectra},  $C(\tau)$ has been computed for the following limiting cases:  
\begin{enumerate}
\item For nonzero ${\bf U}_0$ with  $\mathbf U_0 \cdot \mathbf k \gg \nu(k) k^2$ and $\mathbf U_0 \cdot \mathbf k \gg k \tilde{U}_0(k)$: In this case,
\bea
 C(\tau) & \sim & (\epsilon U_0 \tau)^{2/3},
 \eea
where $\epsilon$ is the dissipation rate that equals the energy flux.  The above $C(\tau)$ yields the following frequency spectrum:
  \be
  E(f)  \sim   (\epsilon U_0)^{2/3} f^{-5/3} ,
  \ee
  which is the prediction of Taylor's frozen-in turbulence hypothesis.  In fact, this idea is often used to test Kolmogorov's spectrum in turbulence experiments~\cite{Tennekes:book}. Recently, Kumar $\&$ Verma (2018) \cite{Kumar:RSOS2018} invoked this scheme to deduce the energy spectrum for  Rayleigh--B\'{e}nard convection in a cube when the large-scale circulation remains steady.   The above spectrum also follows from  Eq.~(\ref{eq:G_k_omega_with_U0}) that yields  the dominant frequency as $\omega = i U_0 k_z$ for this case.  See Appendix~\ref{app:frequency_spectra} for details.  In  Fig.~\ref{fig:freq_spec}(a) we plot $E(f)$ computed using the time series of randomly distributed 50 probes for ${\bf U}_0 = 10\hat{z}$ simulation.  To compute the frequency spectrum $E(f)$, we record the velocity field at  the 50 real space probes. { For the frequency spectrum computation, we run our simulation for a single eddy turnover time with a constant $\Delta t=3\times10^{-5}$, which helps us compute the Fourier transform of the real space data using equispaced FFT.   We sampled the real space data every 33 time  step.}

\item For $\mathbf{U}_0 =0$: In this case,
\bea
 C(\tau) & \sim & \epsilon  \tau
 \eea
  that yields the following frequency spectrum:
    \be
  E(f)  \sim   \epsilon f^{-2}.
  \ee
As shown in the Appendix~\ref{app:frequency_spectra}, the above spectrum is a result of $\exp(-k^2 [\tilde{U}_0( k)]^2 \tau^2)$ (sweeping effect) and $\exp(-\tau/\tau_c)$ (damping term).  The above spectrum can be derived using the dominant frequency relation $\omega = \nu(k) k^2$.  See Appendix~\ref{app:frequency_spectra} for details.  Also see Fig.~\ref{fig:freq_spec}(b) for an illustration of $E(f)$ computed using the time series of randomly distributed 50 probes for the ${\bf U}_0 = 0$ simulation. 
  \end{enumerate}

{ Tennekes $\&$ Lumley (1972) termed the  correlation associated with $f^{-5/3}$ spectrum as Eulerian space-time correlation due to its connection with the mean flow ${\bf U}_0$ that advects the flow, reminiscent of Eulerian picture.  The frequency spectrum $f^{-2}$ is associated with the fluctuating ``mean velocity$"$, hence Tennekes $\&$ Lumley (1972) called the associated correlation as Lagrangian space-time correlation, possibly relating the sweeping effect with random scattering of particles.  Note however that we derived both these spectra in Eulerian hydrodynamics framework.}

{  Recently He \textit{ et al.} (2010) and He $\&$ Tong (2011) combined  the sweeping effect with Taylor's frozen-in turbulence hypothesis in a framework called {\em elliptic approximation}. Due to the sweeping effect, the isocontour lines of the equal-time correlation function measured at two different locations separated by ${\bf r}_E$ is the following elliptical function (deviates from a straight line, a prediction of Taylor's hypothesis):
 \begin{equation}
r_E^2 = r_{Ez}^2 +   |{\bf r}_{E\perp}|^2 =  [r-(U_0+\tilde{U}_{0z})\tau]^2 + [ |\tilde{\bf U}_{0\perp}| \tau]^2.
\label{eq:rE}
\end{equation}  
In the above expression, $\tilde{U}_{0z}$ and $\tilde{\bf U}_{0\perp}$ are the parallel and perpendicular components of the random velocity field along and perpendicular to ${\bf U}_0$.  Refer to  He \textit{ et al.} (2010) and He $\&$ Tong (2011) for further details.   Thus, the sweeping effect, Taylor's frozen-in turbulence hypothesis, and space-time correlation functions are related to each other.  }

We conclude in the next section.

\section{Discussions and Conclusions}
\label{sec:conclusion}

Using numerical simulations, we demonstrate the presence of sweeping effect in hydrodynamic turbulence.   For  zero mean flow (${\bf U}_0=0$), we compute the velocity correlation function $C(k,\tau)$ and show that its magnitude  decays with time-scale $\tau_c \approx 1/(\nu(k) k^2)$, where $\nu(k)$ is the renormalised viscosity.  However, the phase of the correlation function shows a linear increase with $\tau$ till approximately one eddy turnover time; this is attributed to the sweeping of the small scale fluctuations  by the {\em random mean velocity}, ${\bf \tilde{U}}_0$, of the flow.  Thus we demonstrate a clear signature of sweeping effect in hydrodynamic turbulence.   Note that the phase of the correlation function extracts the effects of the sweeping effect by random mean velocity.  Our approach deviates from those of Sanada $\&$ Shanmugasundaram (1992) who only study the absolute of correlation function and argued that $\tau_c \sim k^{-1}$, in line with the predictions of Kraichnan (1964).  On the contrary, our simulations shows that $\tau_c \sim k^{-2/3}$.

For nonzero mean flow (${\bf U}_0=10\hat{z}$), the correlation function exhibits damped oscillations with a frequency of $\omega=U_0 k$ and decay time scale of $1/(\nu(k) k^2)$; the decay time scales for ${\bf U}_0=10\hat{z}$ is same as that for ${\bf U}_0=0$. A careful examination of the  phase of the correlation function also shows additional variations due to the sweeping by the random mean velocity ${\bf \tilde{U}}_0$ of the flow.

For the aforementioned two cases, the frequency spectra of the velocity field measured by the real-space probes are different.  For ${\bf U}_0=0$, $E(f) \sim f^{-2}$, which is related to the Lagrangian space-time correlation, but for ${\bf U}_0=10\hat{z}$, $E(f) \sim f^{-5/3}$, which is the predictions of Taylor's frozen-in turbulence  hypothesis.  We demonstrate these spectra using their respective space-time correlation functions.  Our analysis shows that Taylor's hypothesis is applicable when
\be
U_0 k \gg \nu(k) k^2;~~~U_0 \gg \tilde{U}_0,
\ee
where $\tilde{U}_0$ is random mean velocity, which is responsible for the sweeping effect.

Thus, we provide a systematic demonstration of sweeping effect and Taylor's  frozen-in turbulence hypothesis, and show consistency between the two contrasting phenomena.  We demonstrate the above spectra using numerical simulations.

\section*{Acknowledgements}
We thank Sagar Chakraborty, K. R. Sreenivasan, Robert Rubinstein, Victor Yakhot,  Jayanta K. Bhattacharjee, and Avishek Ranjan for useful discussions and suggestions.  Our numerical simulations were performed on {\em Chaos} clusters of IIT Kanpur, and on Shaheen II of the Supercomputing Laboratory at King Abdullah University of Science and Technology (KAUST) under the project K1052. This work was supported by the research grants PLANEX/PHY/2015239 from Indian Space Research Organisation, India, and  by the  Department of Science and Technology, India (INT/RUS/RSF/P-03) and Russian Science Foundation Russia (RSF-16-41-02012) for the Indo-Russian project.

\appendix
\section{Sweeping effect and renormalization in Eulerian framework}
\label{app:rgcalc}
 
 {
 In this section we extend iterative renormalization group (i-RG) of McComb (1990)
and Zhou (2010) to include the effects of the mean velocity field ${\bf U}_0$.  We show that the renormalized viscosity is independent of ${\bf U}_0$. However, this scheme  fails to capture the sweeping effect.  This issue was first raised by  Kraichnan (1964) in direct interaction approximation (DIA) framework.    Note that the above computations are based on Eulerian framework.  Since the above RG scheme is covered in detail in many papers, here we highlight the changes induced by ${\bf U}_0$. 
 
In  Fourier space, the Navier--Stokes equations in the presence of ${\bf U}_0$  are  (McComb
1990)
\begin{eqnarray}
(-i\omega+  i {\mathbf U_0   \cdot \mathbf k}  +   \nu k^{2} )u_{i}(\hat{{k}})  & = &   -\frac{i}{2}P_{ijm}(\mathbf{k})\int_{\hat{{p}}+\hat{{q}}=\hat{{k}}}d\hat{{p}}\left[u_{j}(\hat{{p}})u_{m}(\hat{{q}})\right] + f_i(\hat{k}) ,\label{eq:udot}  \\
k_i u_i({\bf k}) & = & 0,
\end{eqnarray}
where
\begin{eqnarray}
P_{ijm}(\mathbf{k})  =  k_{j}P_{im}(\mathbf{k})+k_{m}P_{ij}(\mathbf{k}),\\
\label{eq:Pp}
\hat{k} = (\omega, \mathbf k), \hat{p} = (\omega', \mathbf p), \mathrm{and} \, \, \hat{q} = (\omega'', \mathbf q);~~  \hat{k}  = \hat{p}  + \hat{q} . \nonumber
\end{eqnarray}
 We  compute the renormalized viscosity in the presence of a mean velocity $\mathbf U_0$.
In the renormalization process, the wavenumber range $(k_{N},k_{0})$ is divided logarithmically into $N$ shells. The $n$th shell is $(k_{n},k_{n-1})$ where $k_{n}=h^{n}k_{0}\,\,(h<1)$, and $k_N = h^N k_0$. In the first step,  the spectral space is divided in two parts: the shell $(k_{1},k_{0})=k^{>}$, which is to be eliminated, and  $(k_{N},k_{1})=k^{<}$, set of modes to be retained.  The velocity modes in the $k^{>}$ regime are averaged.  The averaging procedure enhances the  viscosity, and the new viscosity is called ``renormalized viscosity$"$.   The process is continued for other shells that leads to larger and larger viscosity.

In i-RG scheme, after $(n+1)$st step, the renormalized equation appears as
 \begin{eqnarray}
 \bigl[ -i\omega && +  i  {\mathbf U_0 \cdot \mathbf k} + ( \nu_{(n)}(k) +\delta\nu_{(n)}(k)) k^2 \bigr] u_{i}^{<}(\hat{k})  = \nonumber \\
&& -\frac{i}{2}P_{ijm}({\textbf{k}}) \int_{\hat{{p}}+\hat{{q}} = \hat{k}} \frac{d \mathbf p d\omega'}{(2\pi)^{d+1}} [u_{j}^{<}(\hat{p})u_{m}^{<}(\hat{k}-\hat{p})] + f^<_i(\hat{k} )
\label{eq:renormalized}
\end{eqnarray}
 with 
\begin{eqnarray}
\delta \nu_{(n)}(\hat{k}) k^2  = \frac{1}{d-1}\int_{\hat{p}+\hat{q}=\hat{k}}^{\Delta}  \frac{d\mathbf p d\omega'}{(2\pi)^{d+1}} [B(k,p,q)G(\hat{q})C(\hat{p})].
\label{eq:renormalized_GC}
\end{eqnarray}
In the above expression,
\begin{equation}
B(k,p,q)=k p [(d-3)z+2z^{3}+(d-1)x y],  \label{eq:Bkpq}
\end{equation}
where $d$ is the space dimensionality,  $x,y,z$ are the direction cosines of ${{\mathbf k}, {\mathbf p}, {\mathbf q}}$, and  $G(\hat{q}), C(\hat{p})$ are respectively Green's and correlation functions that are defined as~\cite{McComb:book:Turbulence,Zhou:PR2010,Verma:PR2004} 
\bea
G(\hat{q}) & = & \frac{1}{-i \omega'' + i {\mathbf U_0 \cdot \mathbf q} + \nu_{(n)}(q) q^2},
\label{eq:Green_fn_RG} \\
C(\hat{p}) & = & \frac{C({\bf p})}{-i \omega' + i {\mathbf U_0 \cdot \mathbf p} + \nu_{(n)}(p) p^2}.
\label{eq:correlation_fn_RG}
\eea

Using  $\omega = \omega' + \omega''$, we obtain
\begin{eqnarray}
 \delta \nu_{(n)}(\omega, k)  k^2   & = &    \frac{1}{d-1} \int_{\hat{p}+\hat{q}=\hat{k}}  \frac{d\mathbf p d\omega'}{(2\pi)^{d+1}}  B(k,p,q)  C(\mathbf p) \nonumber \\ \nonumber &&  \times \frac{1}{\bigl[ -i\omega + i\omega'  + i {\mathbf U_0 \cdot \mathbf q} + \nu_{(n)}(q) q^2 \bigr]  \bigl[-i\omega' + i {\mathbf U_0 \cdot \mathbf p} + \nu_{(n)}(p)p^2 \bigr]} \nonumber \\
 &   = &  \frac{1}{d-1}  \int_{\bf {p + q = k}}^{\Delta}  \frac{d\mathbf p}{(2\pi)^d}
\frac{B(k,p,q)  C(\mathbf p)}{ \bigr[ -i(\omega - {\mathbf U}_0 \cdot {\mathbf k}) + \nu_{(n)}(p) p^2 + \nu_{(n)}(q) q^2 \bigr]}   \nonumber \\
&= &  \frac{1}{d-1}  \int_{\bf {p + q = k}}^{\Delta} \frac{d\mathbf p}{(2\pi)^d}
 \frac{B(k,p,q)  C(\mathbf p)}{   \nu_{(n)}(p) p^2 + \nu_{(n)}(q) q^2 }.
\label{eq:delta_nu2}
\end{eqnarray}
Note that  $ \omega - {\mathbf U}_0 \cdot {\mathbf k}= \omega_D$ is the Doppler-shifted frequency in the moving frame, where the frequency of the signal  is reduced.  It is analogous to the reduction of frequency of the sound wave in a moving train when the train moves away from the source.  For ${\bf U}_0=0$, it is customary to assume that $\omega \rightarrow 0$ since we focus on dynamics at large time scales~\cite{McComb:book:Turbulence,Zhou:PR2010,McComb:book:new}.  The corresponding assumption for ${\bf U}_0\ne 0$ is to set $\omega_D \rightarrow 0$ because $\omega_D$ is the effective frequency of the large scale modes in the moving frame.  The approximation   $ \omega \rightarrow \omega_D$ essentially takes away the effect of Galilean transformation and provides inherent turbulence properties.  Note that in Taylor's frozen-in turbulence hypothesis, $\omega = {\bf U_0 \cdot k}$ that yields $\omega_D = 0$~\cite{Tennekes:book}.

Equation~(\ref{eq:delta_nu2}) indicates that the correction in viscosity, $ \delta \nu_{(n)}$, is independent of ${\bf U}_0$. After this step, the derivation of renormalized viscosity with and without ${\bf U}_0$ are identical. 

Equation~(\ref{eq:delta_nu2}) however does not include any sweeping effect, which is a serious limitation  of Eulerian field theory, as pointed out by  Kraichnan (1964) in direct interaction approximation (DIA) framework.   Kraichnan (1965) then formulated Lagrangian-history closure approximation for turbulence  and showed consistency with Kolmogorov's spectrum (also see Leslie 1973).  Effectively, a consistent theory needs to include a term of the form $i\tilde{\bf U}_0 \cdot {\bf q}$  in the denominator of Eq.~(\ref{eq:Green_fn_RG}).  A procedure adopted by Verma (1999) \cite{Verma:PP1999} for ``mean magnetic field$"$ renormalization in magnetohydrodynamic turbulence may come out to be  handy for such computations, which may be attempted in future.   
  
\section{Computation of spatio-temporal correlations and frequency spectra of turbulent flow }
\label{app:frequency_spectra}

Using the normalized correlation function of Eq.~(\ref{eq:Rk_tau_with_U0}),  we derive the following spatio-temporal correlation function:
\bea
 C({\bf r}, \tau) & = & \int d{\bf k} C({\bf k}) \exp(-\nu(k) k^2 \tau-i{\bf U_0 \cdot k} \tau)  \exp(-i {\bf k} \cdot \tilde{\bf U}_0({\bf k}) \tau)  \exp(i {\bf k} \cdot {\bf r}).
 \label{eq:corr_U0}
\eea 
We time average $\tilde{U}_0$ over random ensemble~\cite{Kraichnan:PF1964Eulerian, Wilczek:PRE2012} that yields
\bea
 C({\bf r}, \tau) & = & \int d{\bf k} C({\bf k}) \exp(-\nu(k) k^2 \tau - i{\bf U_0 \cdot k} \tau) \langle \exp(-i c k \tilde{U}_0( k)\tau)  \rangle \exp(i {\bf k} \cdot {\bf r}) \nonumber \\
 & = & \int d{\bf k} C({\bf k}) \exp(-\tau/\tau_c - i{\bf U_0 \cdot k} \tau )  \exp(-k^2 [\tilde{U}_0( k)]^2 \tau^2)  \exp(i {\bf k} \cdot {\bf r}).
\eea 
In addition, we  set ${\bf r}=0$ to compute the temporal correlation at  a single point. 

In the above integral, following Pope (2000) \cite{Pope:book}, we replace the isotropic and homogeneous $C({\bf k})$ with 
\be
C({\bf k})  = \frac{E(k)}{4\pi k^2} = \frac{f_L(kL) f_\eta(k\eta) K_\mathrm{Ko} \epsilon^{2/3} k^{-5/3}}{4\pi k^2},
\label{eq:Ek_Pope}
\ee
where $\epsilon$ is the energy dissipation rate, which is same as the energy flux, and
\begin{eqnarray}
f_L(kL) & = & \left( \frac{kL}{[(kL)^2 + c_L]^{1/2}} \right)^{5/3+p_0}, 
\label{eq:fL} \\
f_\eta(k\eta) & = & \exp \left[ -\beta \left\{ [ (k \eta)^4 + c_\eta^4 ]^{1/4}   - c_\eta \right\} \right],
\label{eq:feta}
\end{eqnarray}
with $c_L, c_\eta, p_0, \beta$ as constants, and $L$ as the large length scale of the system.   We also substitute $\tau_c(k) = 1/(\nu(k) k^2) = \epsilon^{-1/3}k^{-2/3}$ and $\tilde{U}_0(k) = \epsilon^{1/3}k^{-1/3}$ (from dimensional analysis).  We ignore the coefficients in front of these quantities for brevity.   After the above substitutions, we obtain
\bea
 C(\tau) & = &  K_\mathrm{Ko} \epsilon^{2/3} \int dk  k^{-5/3} f_L(kL) f_\eta(k\eta) \exp(-i{\bf U_0 \cdot k} \tau)  \times  \nonumber \\
 	&&  \exp(-\epsilon^{1/3}k^{2/3}\tau )  \exp(-\epsilon^{2/3}k^{4/3}\tau^2).
	\label{eq:C_tau_appendix}
\eea 

The above form of $C(\tau)$ is valid for any ${\bf U}_0$.  The above integral is too complex, hence we perform asymptotic analysis in two limiting cases that are described below.

\subsection{For $\mathbf U_0 \cdot \mathbf k \gg \nu(k) k^2$ and $\mathbf U_0 \cdot \mathbf k \gg k \tilde{U}_0(k)$}

For this case $U_0$ dominates other velocity scales,  hence we take $\tau \sim 1/(U_0 k)$ as the dominant time scale.  For simplification, we  make a change of variable, $\tilde{k} =   U_0 k \tau$.  In addition, we choose the $z$ axis to be along the direction of ${\bf U}_0$.  Under these simplifications, the integral becomes
\bea
 C(\tau) & \approx &  K_\mathrm{Ko}  (\epsilon U_0 \tau)^{2/3}  \int d{\tilde{k}} \tilde{k}^{-5/3} f_L(\tilde{k} (L/U_0\tau)) f_\eta( \tilde{k} (\eta / U_0 \tau)  \frac{\sin(U_0 k \tau)}{U_0 k \tau} \times   \nonumber \\
 &&  \exp[- \tilde{k}^{2/3} (U/U_0)^{2/3} (\tau/T)^{1/3} - \tilde{k}^{4/3} (U/U_0)^{4/3} (\tau/T)^{2/3} ].
\eea 
We focus on $\tau$ in the inertial range, hence $L/U_0\tau \gg 1$ and $\eta/ U_0 \tau \ll 1$,  consequently,  $f_L(\tilde{k} (L/U_0\tau)) \approx 1$, and $f_\eta( \tilde{k} (\eta / U_0 \tau) \approx 1$.     Therefore,
\bea
 C(\tau) & \approx & K_\mathrm{Ko} (\epsilon U_0 \tau)^{2/3}   \int d{\tilde{k}} \tilde{k}^{-5/3} \frac{\sin \tilde{k}}{\tilde{k} }  \exp[- \tilde{k}^{2/3} (U/U_0)^{2/3} (\tau/T)^{1/3} - \tilde{k}^{4/3} (U/U_0)^{4/3} (\tau/T)^{2/3} ] \nonumber \\ 
 & \approx & B  K_\mathrm{Ko}(\epsilon U_0 \tau)^{2/3},
\eea 
where $B$ is the value of the nondimensional integral.  The Fourier transform of the above $ C(\tau)$ yields the following frequency spectrum:
\bea
E(f) & \approx & \int C(\tau) \exp(i 2\pi f \tau) d\tau = \int  B K_\mathrm{Ko}(\epsilon U_0 \tau)^{2/3} \exp(i 2\pi f \tau) d\tau  \nonumber \\
& \sim  &(\epsilon U_0)^{2/3} f^{-5/3}.
\eea
The above frequency spectrum is the prediction of Taylor's frozen-in turbulence hypothesis. 

\subsection{For $\mathbf{U}_0=0$}
We set $\mathbf{U}_0  =0$ in Eq.~(\ref{eq:C_tau_appendix}).  In the resulting equation, both the remaining exponential terms (the damping and sweeping effect terms) have the following time scale:
\be
\tau(k) \sim 1/(k u_k) \sim \epsilon^{-1/3} k^{-2/3}.
\ee
Hence, for computing the integral $C(\tau)$, we make a change of variable:
\be
k = \tilde{k} \epsilon^{-1/2} \tau^{-3/2}
\ee
that transforms the integral to
\bea
 C(\tau) & \approx &  K_\mathrm{Ko} \epsilon  \tau \int d{\tilde{k}} \tilde{k}^{-5/3} f_L(\tilde{k} (L/U\tau)^{3/2}) f_\eta( \tilde{k} (\tau_d/ \tau)^{3/2})  \exp(- \tilde{k}^{2/3} - \tilde{k}^{4/3} ), \label{eq:C_tau_U_zero}
\eea 
where $U$ is the large-scale velocity, and $\tau_d$ is the dissipative time scale.  We focus on $\tau$ in  the inertial range, hence $L/U\tau \gg 1$ and $\tau_d/ \tau \ll 1$.  Therefore, using Eqs.~(\ref{eq:fL}),~(\ref{eq:feta}), we deduce that $f_L(\tilde{k} (L/U\tau)^{3/2}) \approx 1$ and $f_\eta( \tilde{k} (\tau_d/ \tau)^{3/2}) \approx 1$.     Therefore,
\bea
 C(\tau) \approx  K_\mathrm{Ko} \epsilon  \tau \int d{\tilde{k}} \tilde{k}^{-5/3} \exp(- \tilde{k}^{2/3} - \tilde{k}^{4/3} ) \approx A K_\mathrm{Ko} \epsilon  \tau,
 \label{eq:C_tau_U_zero_2}
\eea 
where $A$ is the value of the integral of Eq.~(\ref{eq:C_tau_U_zero_2}).  The Fourier transform of $ C(\tau)$ yields the following frequency spectrum:
\bea
C(f) & = & \int C(\tau) \exp(i 2\pi f \tau) d\tau = A K_\mathrm{Ko} \epsilon \int \tau  \exp(i 2\pi f  \tau) d\tau  \sim  \epsilon f^{-2}.
\label{eq:Eomega_minus2}
\eea
 Thus, the damping and sweeping terms yield  frequency spectrum $E(f) \sim f^{-2}$.  

We could also derive the above frequency spectra using scaling arguments~\cite{Landau:book:Fluid}.  From Eq.~(\ref{eq:G_k_omega_with_U0}), we obtain the dominant frequency as
\be
\omega = \mathbf U_0 \cdot \mathbf k +c k \tilde{U}_0(k)  -i \nu(k) k^2.
\ee
  When $\mathbf U_0 \cdot \mathbf k \gg \nu(k) k^2$ and $\mathbf U_0 \cdot \mathbf k \gg k \tilde{U}_0(k)$, we obtain $\omega = U_0 k_z$.  Therefore, using  the formula for one-dimensional spectrum $E(k) = K_\mathrm{Ko} \Pi^{2/3} k^{-5/3}$, and $\omega = 2 \pi f$, we obtain
\begin{eqnarray}
E(f) & = &  E(k) \frac{dk}{df}  \sim  (\epsilon U_0)^{2/3} f^{-5/3}.
\label{eq:f5by3scaling}
\end{eqnarray}
 On the contrary, when $\mathbf U_0 \cdot \mathbf k \ll\nu(k) k^2$ (for zero or small $U_0$), we obtain  $
\omega \approx  \nu(k) k^2 = \nu_* \sqrt{K_\mathrm{Ko}} \Pi^{1/3} k^{2/3}$ 
and hence,  
\begin{eqnarray}
E(f) & = & E(k) \frac{dk}{df} \sim  \Pi f^{-2},
\label{eq:f2scaling}
\end{eqnarray}
consistent with the formulas derived earlier.  
}



\nocite{}

\newpage

\setlength{\tabcolsep}{20pt}
\begin{table}
\begin{center}
\begin{tabular}{c  c c  c c}
$\rm Grid$  &${\bf U}_0$ & $\nu$  &$\mathrm{Re}$ & $k_{\rm max}\eta$ \\[1 mm]
$512^3$ & $0$ & $10^{-3}$ & $5.7 \times 10^3$ &$2.5$\\
$512^3$ & $10\hat{z}$ & $10^{-3}$ & $5.7 \times 10^3$ &$2.5$\\
$1024^3$ & $0$ & $4 \times 10^{-4}$ & $1.3 \times 10^4$ &$2.5$\\
\end{tabular}
\caption{ Parameters of our direct numerical simulations (DNS): Grid resolution; Mean velocity ${\bf U}_0$; Kinematic viscosity $\nu$; Reynolds number $\mathrm{Re} = u_\mathrm{rms}L/\nu$; and $k_{\rm max}\eta$, where { $k_{\rm max}=N/2$} is the maximum wavenumber, and $\eta$ is the Kolmogorov length.}
\label{table:simulation_details}
\end{center}
\end{table}

\end{document}